\documentclass[10pt,conference]{IEEEtran}

\IEEEoverridecommandlockouts
\usepackage{microtype}
\usepackage{cite}
\usepackage{graphicx}
\usepackage{booktabs}
\usepackage{siunitx}
\usepackage{paralist}
\usepackage{multirow}
\usepackage{array}
\usepackage{graphicx}
\usepackage{fontawesome}
\usepackage[para]{footmisc}
\usepackage{makecell}
\usepackage{balance}
\usepackage{hyperref}
\usepackage[hyphenbreaks]{breakurl}

\usepackage{xcolor}
\usepackage{colortbl}
\usepackage{wrapfig}
\usepackage{subfig}

\newcommand*{\belowrulesepcolor}[1]{%
  \noalign{%
    \kern-\belowrulesep
    \begingroup
      \color{#1}%
      \hrule height\belowrulesep
    \endgroup
  }%
}
\newcommand*{\aboverulesepcolor}[1]{%
  \noalign{%
    \begingroup
      \color{#1}%
      \hrule height\aboverulesep
    \endgroup
    \kern-\aboverulesep
  }%
}

\usepackage[noend]{algorithmic}
\usepackage{algorithm}

\usepackage{syntax}
\renewcommand{\litleft}{\sf \begin{tikz}[baseline=(X.base)]\node [draw=gray!40,fill=gray!0,semithick,rectangle,inner sep=1pt, minimum size=1em, outer sep=0pt, rounded corners=2pt] (X) }
\renewcommand{\litright}{;\end{tikz} \normalfont}

\usepackage{enumitem}
\usepackage{xspace}
\usepackage{xargs}
\usepackage{amsthm}
\usepackage{adjustbox}
\usepackage{microtype}
\usepackage{pifont}
\newcommand\definetool[2]{\newcommand{#1}{{\textsc{#2}}\xspace}}
\definetool{\scratch}{Scratch}
\definetool{\drscratch}{Dr. Scratch}
\definetool{\leila}{LeILa}
\definetool{\whisker}{Whisker}
\definetool{\litterbox}{LitterBox}
\definetool{\bastet}{Bastet}
\definetool{\hairball}{Hairball}
\definetool{\qualityhound}{QualityHound}
\definetool{\scratchblocks}{scratchblocks}

\newcommand{\blockleft}{\begin{mbox}\sf\begin{tikz}[baseline=(X.base)]\node[draw=black!60,fill=black!3,semithick,rectangle,inner sep=1pt, minimum size=1em, outer sep=0pt, rounded corners=1pt] (X)}%
\newcommand{\blockright}{;\end{tikz}\normalfont\end{mbox}}%
\newcommand{\scratchblock}[1]{\blockleft{\small\textsf{#1}}\blockright}

\usepackage{multirow}
\usepackage{array}
\newlength{\defaulttabcolsep}
\definecolor{OliveGreen}{rgb}{0,0.6,0}
\usepackage[colorinlistoftodos,prependcaption,textsize=tiny]{todonotes}
\newcommandx{\missing}[2][1=]{\todo[linecolor=red,backgroundcolor=red!25,bordercolor=red,#1]{#2}}
\newcommandx{\unsure}[2][1=]{\todo[linecolor=orange,backgroundcolor=orange!25,bordercolor=orange,#1]{#2}}
\newcommandx{\change}[2][1=]{\todo[linecolor=blue,backgroundcolor=blue!25,bordercolor=blue,#1]{#2}}
\newcommandx{\info}[2][1=]{\todo[linecolor=OliveGreen,backgroundcolor=OliveGreen!25,bordercolor=OliveGreen,#1]{#2}}

\usepackage{fbox}
\usepackage[formats]{listings}
\definecolor{lightgray}{rgb}{.9,.9,.9}
\definecolor{darkgray}{rgb}{.4,.4,.4}
\definecolor{purple}{rgb}{0.65, 0.12, 0.82}
\definecolor{rltred}{rgb}{0.5,0,0}
\definecolor{rltgreen}{rgb}{0,0.5,0}
\definecolor{rltblue}{rgb}{0,0,0.5}
\definecolor{DarkGreen}{rgb}{0.00,0.60,0.00}
\definecolor{ScarletRed}{rgb}{0.80,0.00,0.00}
\definecolor{blizzardblue}{rgb}{0.67, 0.9, 0.93}
\definecolor{green-yellow}{rgb}{0.68, 1.0, 0.18}

\definecolor{dkgreen}{rgb}{0,0.6,0}
\definecolor{gray}{rgb}{0.5,0.5,0.5}
\definecolor{mauve}{rgb}{0.58,0,0.82}
\definecolor{lightgrey}{rgb}{0.90,0.90,0.90}

\definecolor{grey}{gray}{0.75}
\definecolor{light-gray}{gray}{0.80}
\usepackage[most]{tcolorbox}

\newtcblisting{commandshell}{colback=black,colupper=white,colframe=yellow!75!black,boxsep=0mm,left=1mm,listing only,listing options={language=sh,basicstyle=\scriptsize\ttfamily},every listing line={\textcolor{red}{\scriptsize\ttfamily\bfseries \$ }}}

\begin{document}

\title{LitterBox: A Linter for Scratch Programs}

\author{\IEEEauthorblockN{Gordon Fraser, Ute Heuer, Nina K\"{o}rber, Florian Oberm\"{u}ller, Ewald Wasmeier\thanks{Authors listed in alphabetical order}}
 \IEEEauthorblockA{
 \textit{University of Passau}\\
 Passau, Germany}
 }


\newcommand{\numnoremix}{74,830\xspace}
\newcommand{\numbugs}{109,951\xspace}
\newcommand{\numclassified}{250\xspace}
\newcommand{\numfalsepositives}{32\xspace}

\maketitle

\begin{abstract}
Creating programs with block-based programming languages like \scratch is easy and fun. Block-based programs can nevertheless contain bugs, in particular when learners have misconceptions about programming. Even when they do not, \scratch code is often of low quality and contains code smells, further inhibiting understanding, reuse, and fun.
To address this problem, in this paper we introduce \litterbox, a linter for \scratch programs. Given a program or its public project ID, \litterbox checks the program against patterns of known bugs and code smells. For each issue identified, \litterbox provides not only the location in the code, but also a helpful explanation of the underlying reason and possible misconceptions.
Learners can access \litterbox through an easy to use web interface with visual information about the errors in the block-code, while for researchers \litterbox provides a general, open source, and extensible framework for static analysis of \scratch programs. 
\end{abstract}


\section{Introduction}

The \scratch~\cite{maloney2010} block-based programming language is
tremendously popular amongst teachers and programming novices. While it is easy
to create games and animations using \scratch~\cite{scratch-for-all}, programs
may nevertheless contain bugs. This is particularly the case because \scratch
programmers are usually young learners who might not know, or have
misconceptions about, programming concepts. Even when programs suitably satisfy
the behavior intended by the programmer, the quality of \scratch code has been
reported to be low and riddled with code smells~\cite{hermans2016a,
techapalokul2017a}. This may have serious implications on learning outcomes,
the general understandability of \scratch programs, and the overall
enjoyment of programming. It can also severely inhibit teachers who may face the
daunting task of debugging a potentially large and diverse set of student
solutions at the same time.

\begin{figure}[t]
    \centering
	\subfloat[\label{fig:example-intro}Erroneous \scratch script.]{\includegraphics[width=0.4\columnwidth]{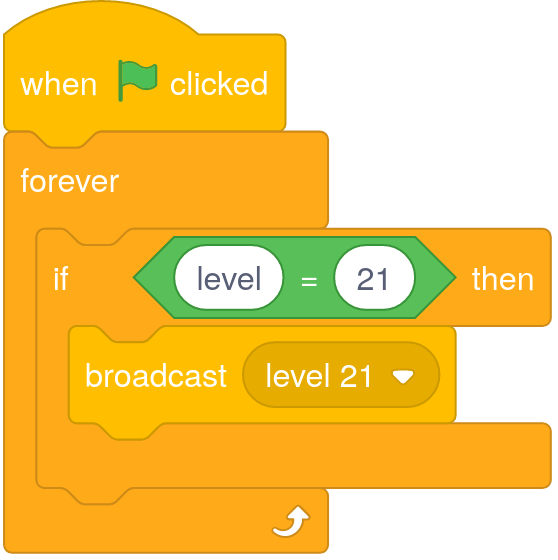}}\hfill
	\subfloat[\label{fig:example-intro-pattern}Bug pattern in the syntax tree.]{\includegraphics[width=0.5\columnwidth]{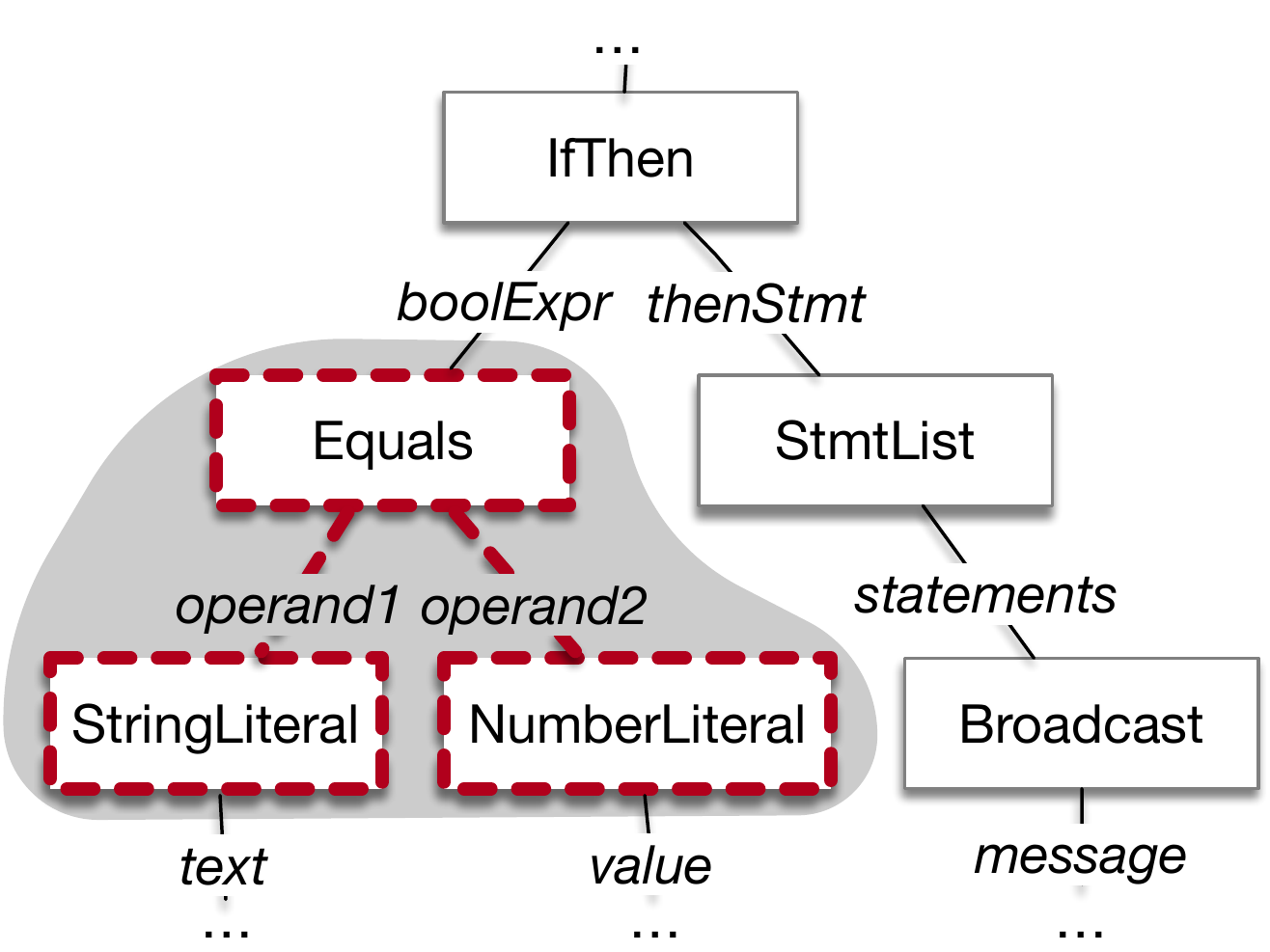}}\\
	\subfloat[\label{fig:example-intro-hint}Hint provided by \litterbox.]{\includegraphics[width=\columnwidth]{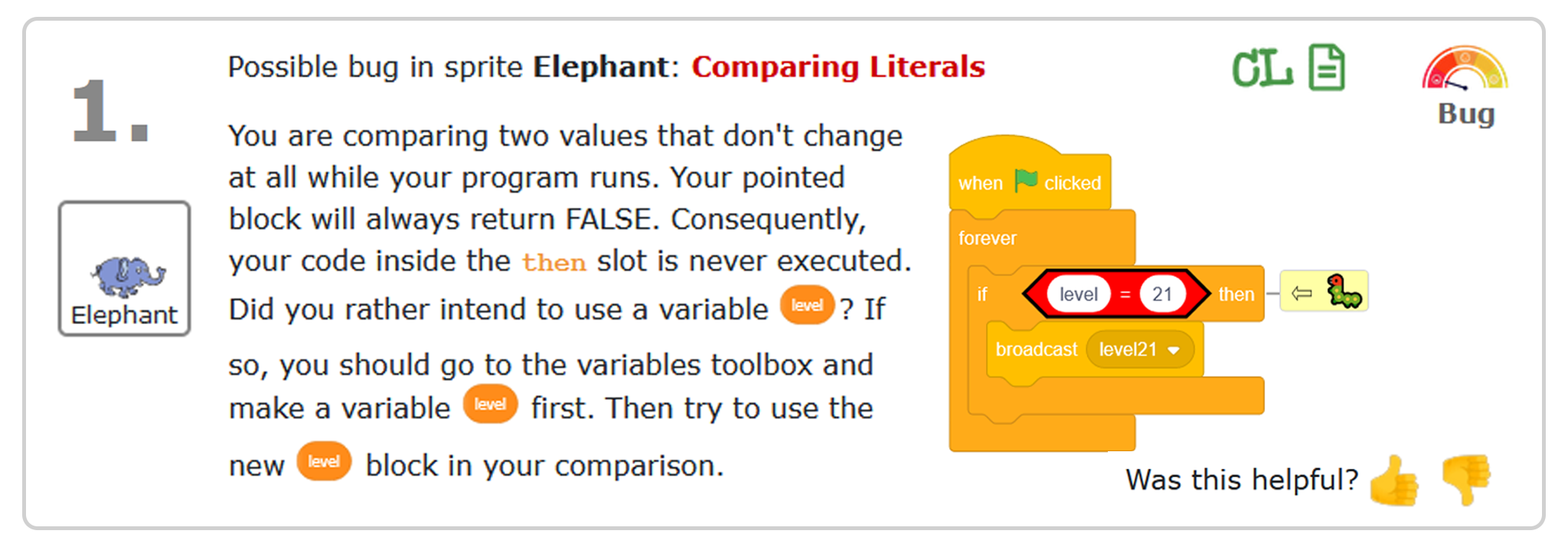}}
    \caption{Example bug pattern: Instead of comparing the variable \emph{level} with the value 21, this script in the ``Elephant'' sprite compares the literal ``level'' with 21---a comparison that will never evaluate to true.}
\end{figure}

In order to address this problem, we introduce \litterbox, a linter for \scratch programs. \litterbox is built on the observation that, even though there are abundant ways to produce bugs, many of them result from similar misconceptions and manifest in common \emph{patterns} of bugs.
As an example, consider the script shown in Figure~\ref{fig:example-intro}, which was written by a \scratch user with the intent to continuously check whether level 21 has been reached in a game, and if so to broadcast a corresponding message. The program contains a bug that is typical of learners who have not yet fully comprehended the concept of variables: Instead of comparing the value of the variable \emph{level} with 21, the literal ``level'' is compared with the number 21. Obviously, this if-condition will never evaluate to true. Such a comparison between literals can easily be described as a pattern on the abstract syntax tree of \scratch programs, and bugs matching this pattern occur frequently: In a recent study, we found 4,939 instances of this bug pattern in a dataset of 74,830 publicly shared \scratch projects~\cite{fraedrich2020}.

\litterbox identifies bugs like this by statically analyzing the source code of \scratch programs, and applying a catalogue of finders for different bug patterns and code smells on the abstract syntax tree (Figure~\ref{fig:example-intro-pattern}) and control flow graph. For each issue identified, \scratch provides a visual summary of the relevant blocks together with an explanation of the issue, as well as possible underlying misconceptions. Figure~\ref{fig:example-intro-hint} illustrates this for the \emph{comparing literals} bug seen in Figure~\ref{fig:example-intro}.

\litterbox is developed as an open source Java project. On the command line, \litterbox can be used to check individual \scratch projects, collections of them, or for downloading projects from the \scratch website given their project IDs, and producing reports in configurable output formats. In addition to the command line interface, we provide a web interface which makes \litterbox easy to use for learners as well as for teachers having to assess multiple projects.


\section{Code Quality Issues in \scratch}

\subsection{The \scratch Programming Language}

\scratch~\cite{maloney2010} is popular amongst teachers and programming novices
for several reasons:
It is block-based, which means that programs are assembled by visually arranging blocks representing program statements and expressions. Blocks come in different shapes to visually demonstrate which combinations are valid. For example, boolean expressions are represented as pointed blocks, and other expressions in rounded blocks, and wherever such blocks can be inserted, there are holes in the appropriate shapes. Statements `snap' together like Lego bricks, such that it is only possible to create programs of syntactically valid combinations of blocks.
The blocks available are displayed in a toolbox, supporting recognition over recall. 
While \scratch blocks contain many standard programming language constructs such as conditions and loops, many statements represent high-level actions of the \emph{sprites} interacting on a \emph{stage}. This makes it easy to quickly arrange blocks in a way that results in fun and game-like programs.

While the tinkering-approach to programming is initially encouraging for learners, it has been observed that \scratch programmers tend to 
develop certain negative habits while coding~\cite{meerbaum2011habits}. Multiple studies have demonstrated that issues in code quality are prevalent in \scratch~\cite{aivaloglou2016kids,hermans2016a,techapalokul2017a,robles2017software} and have a negative impact on code understanding~\cite{hermans2016b}. 
\litterbox therefore aims to identify recurring issues in \scratch code~\cite{fraedrich2020}. In the following, we discuss different categories of issues that occur in \scratch programs.

\subsection{Syntax Errors in \scratch}

Although the block-based nature of \scratch is intended to prevent syntax errors, it can only achieve this to a certain degree. Even though it is \emph{usually} not possible to combine blocks in invalid ways, there are exceptions. For example, all `reporter'-blocks have the same shape and so it is possible to use a reporter block for a costume name at a place where a color is expected. The notion of custom blocks gives rise to possible bugs, as \scratch does not validate the signature of these blocks; for example it is possible to have multiple parameters with the same name, or to use parameter blocks outside their scope. Many different types of syntax errors can arise from reuse and modification, for example when custom blocks are deleted, uses of their parameters may still remain in the program. Finally, incomplete programs can also give rise to bugs; for example, omitting a condition in a \scratchblock{repeat until} loop will lead to an infinite loop.

\subsection{\scratch-specific Bugs}

The program scenario of a stage with multiple sprites interacting on it gives rise to bugs specific to this setting. For example, programs may contain event handlers for scene-related events (e.g., a change of the backdrop) without these events ever being triggered. The ability to use \scratch in a Logo-like mode, where sprites draw lines when moving, gives rise to various types of bugs when omitting to enable or disable the `pen'-mode.

\subsection{Patterns of General Bugs}

Besides \scratch-specific bugs, there are countless ways for producing mistakes that may occur in any programming language. For example, it is possible to produce infinite recursions with custom blocks or message passing, there may be control flow anomalies, and there may be data flow issues such as lack of initialization. Common issues include omissions of \scratchblock{broadcast} or \scratchblock{clone} statements for which handlers are available, or vice versa, omissions of the handlers for events that are produced. There is no end to the creativity with which bugs can be produced in \scratch.

\subsection{Code Smells}

Even when \scratch programs do not contain bugs, the code may nevertheless be of mediocre quality. The concept of \emph{code smells} describes different concrete quality problems in code which is not wrong per-se, but its quality issues reduce understandability and maintainability, and thus increase the chances that bugs are introduced later on. Common code smells known from regular programming languages often also apply to \scratch programs (e.g., spaghetti code, code duplication).


\section{\litterbox: Static Analysis for \scratch}


\begin{table}[tbh!]
        \caption{\label{tab:finders}Issue finders implemented in \litterbox 1.5.}
\resizebox{\columnwidth}{!}{%
\renewcommand{\arraystretch}{1.1}
 \begin{tabular}{lp{5.2cm}}
        \toprule
\belowrulesepcolor{gray!25}
\rowcolor{gray!25}Syntax Errors		& \\
\aboverulesepcolor{gray!25}
        \midrule		
Ambiguous Custom Block Signature  & Several custom blocks have identical hats\\
Ambiguous Parameter Name  & Custom block parameters with identical name\\
Call Without Definition  & Non existing custom block is called\\
Expression As Touching Or Color  & Reporter is used in color or object spot\\
Illegal Parameter Refactor  & String parameter is used in bool condition\\
Missing Termination Condition  & Repeat until without condition\\
Missing Wait-Until Condition  & Wait until without condition\\
Orphaned Parameter  & Parameter is not defined anymore\\
Parameter Out Of Scope  & Parameter outside custom block\\
        \midrule
\belowrulesepcolor{gray!25}
\rowcolor{gray!25}\scratch-specific Bugs		& \\
\aboverulesepcolor{gray!25}
        \midrule
Missing Backdrop Switch  & Backdrop switch event never triggered\\
Missing Erase All  &  Pen lines are not erased\\
Missing Pen Down  &  Pen is up but never down\\
Missing Pen Up  &   Pen is down but never up\\
Missing Resource & Used costume, sound or background is missing\\
Stuttering Movement  &  Annoying typematic delay\\
        \midrule
\belowrulesepcolor{gray!25}
\rowcolor{gray!25}General Bugs		& \\
\aboverulesepcolor{gray!25}
        \midrule
Blocking If-Else & Terminates in both paths, code after if-else \\ 
Comparing Literals  & Strings/Numbers are compared directly\\
Custom Block With Forever  &  Blocks after custom block never execute\\
Custom Block With Termination  & Blocks after custom block never execute\\
Delete Clone After Broadcast & Clone is deleted immediately after broadcast\\
Endless Recursion  & Custom block or script calls itself without termination\\
Forever Inside Loop  & Outer loop is never executed \\
Inappropriate Hatblock & Greenflag handler in script with delete clone \\
Interrupted Loop Sensing & Block that takes time interrupts continuous sensing\\
Message Never Received  & Broadcast does not trigger handler\\
Message Never Sent  & Broadcast for handler is never sent\\
Missing Ask  & Answer is used without ask\\
Missing Clone Call  & Clone event is never called\\
Missing Clone Initialization  & Clone handler is not used\\
Missing Initialization  & Sprite is not initialized \\
Missing Loop Sensing  & Condition is checked only a single time\\
No Working Scripts  & Only empty scripts and code lying around; no handler is connected to any blocks\\
Position Equals Check  & Positions are compared exactly\\
Recursive Cloning  & Clones clone themselves without termination\\
Stop after Say & Script stopped immediately after say \\
Terminated Loop  & Loop is stopped during first iteration\\
Type Error & Incompatible blocks are compared\\
Variable As Literal  & Variable name instead of reporter\\
        \midrule
\belowrulesepcolor{gray!25}
\rowcolor{gray!25}Code Smells		& \\
\aboverulesepcolor{gray!25}
        \midrule
Busy Waiting & Constantly checking to stop script\\
Cloned Code & Code clones of types 1-3\\
Code Lying Around  & Loose blocks without handler\\
Double If  & Consecutive if with same condition\\
Duplicate Sprite  & Two sprites are exact duplicates\\
Duplicated Script  & Two scripts in a sprite are exact duplicates\\
Empty Control Body  & C-block without sub stack\\
Empty Custom Block  & Custom block without body\\
Empty Project  & Project without sprites\\
Empty Script  & Handler without body\\
Empty Sprite  & Sprite without scripts\\
Long Script  & Script longer than 12 blocks\\
Message Naming  & Message with uncommunicative name\\
Middle Man  & Broadcast reception sends next broadcast; custom block calls next custom block \\
Multi Attribute Modification  & Variable is changed multiple times in a row\\
Nested Loops  & Loops without other blocks stacked in-between\\
Same Variable Different Sprite  & Same variable name in multiple sprites\\
Sequential Actions  & Sequence of repeated blocks instead of loop\\
Sprite Naming  &  Sprite with uncommunicative name\\
Unnecessary If After Until & If checks same condition that terminated until\\
Unnecessary Loop & Loop that runs never or one time\\
Unused Custom Block  & Custom block is never called\\
Unused Parameter & Parameter is defined but not used \\
Unused Variable  & Variable is never used\\
Variable Initialization Race  &  Variable is initialized with different values in scripts with same handler\\
        \bottomrule
        \end{tabular}%
        }
		\vspace{-2em}
\end{table}

\subsection{Main Features of \litterbox}

\noindent\textbf{Bug finders.}
The central feature of \litterbox is to find issues in \scratch programs based on bug patterns and code smells. Table~\ref{tab:finders} provides a list of all issue finders implemented in \litterbox 1.5. Extending \litterbox with new issue finders is straightforward, so this list is growing continuously.

\noindent\textbf{Code metrics.}
Besides checking for bugs, \litterbox can extract different metrics on \scratch programs, such as the numbers of blocks, scripts, and sprites, or the overall weighted mean complexity of a program.

\noindent\textbf{Code translation.}
\litterbox can translate \scratch programs to LeILA (Learners' Intermediate Language), the intermediate language used by the model checker \bastet~\cite{stahlbauer2020}, and to the
\scratchblocks\footnote{\url{http://scratchblocks.github.io}, last accessed September 23, 2020.} format.

\noindent\textbf{Output formats.}
\litterbox can produce output in different formats: Besides basic information on the console, \litterbox can produce data files in CSV format, which is useful for researchers conducting analyses on datasets of \scratch programs. \litterbox produces a custom JSON report that contains detailed information about the issues found and their descriptions, which serves as a basis for displaying results to learners. \litterbox can export versions of the \scratch project in which all blocks associated with issues are annotated with comments explaining the issues, and giving hints on how the user could try to fix it.

\noindent\textbf{Mining/downloading.}
\litterbox can analyze individual local files as well as folders containing multiple
projects to check. Alternatively, \litterbox can also
handle individual project IDs or lists of IDs, and will then download these projects from the
\scratch servers before analyzing them.

\noindent\textbf{Language support.} 
\litterbox supports internationalization and currently provides output in English, German, and Spanish.

\subsection{Analysis Engine}

\litterbox supports the analysis of projects in the most recent version of
\scratch (3.0). \scratch projects are saved as \texttt{.sb3} files, which are
zip-archives containing the code of the project as a JSON file and all the
assets such as sounds or costumes. \litterbox can process both, \texttt{.sb3} and JSON
files.
A project is parsed to an abstract syntax tree (AST) in \litterbox, in
which there is a distinct class for each type of block in \scratch. Stack
blocks are represented as \emph{Statements} (\emph{Stmt}), reporter blocks
(including variables, lists and parameters of custom blocks) as \emph{Expressions}. \litterbox also has
wrapper classes to reflect the non typed structure of \scratch programs (e.g.,
with a wrapper a \emph{String Expression} can be put into a \emph{Stmt} expecting
a \emph{Number Expression}, for example when the \scratchblock{username} block is used as parameter of the \scratchblock{move~steps} block).

A visitor pattern is used to traverse this tree. So called \emph{issue finders}, which check for bug patterns or code smells, are implemented as visitors to find idioms of
\scratch blocks. Furthermore, \litterbox creates a control flow graph from the AST which is used in some of the finders, e.g., to
report the missing initialization of variables or attributes of sprites.


\section{Using \litterbox}

\subsection{Command Line Usage}

\litterbox can be accessed on the command line using an executable jar-file.
An overview of all possible command line options can be displayed using the \texttt{--help} option:

\begin{commandshell}
java -jar Litterbox-1.5.jar --help
\end{commandshell}

The main modes of operation are \texttt{--check} to apply checks on a \scratch project, \texttt{--stats} to produce code metrics, and \texttt{--leila} to translate a program to LeILA. For example, to check a project stored on the local hard drive (as either \texttt{.sb3} file or just the \texttt{project.json} file containing the code) with all default checkers on would use the following command:

\begin{commandshell}
java -jar Litterbox-1.5.jar --check --path <filename>
\end{commandshell}
	
Instead of a single file, one can also provide a folder, in which case \litterbox will check all \scratch projects contained in that folder. If the project is not available as a downloaded file yet, one can let \litterbox do this as well:

\begin{commandshell}
java -jar Litterbox-1.5.jar --check --projectid <id>
\end{commandshell}

If no output directory is specified, \litterbox will place the file in a temporary directory.

There are command line options to select the output format. By default, \litterbox will only produce output on the console, but providing an output file name using \texttt{--output=<filename>} will produce a report. The format of this report is deduced from the filename (i.e., \texttt{.csv} or \texttt{.json}). Other options, such as whether to ignore unconnected blocks (i.e., scripts that have no hat-block), which checks to apply, and many other options using parameters can be displayed using the \texttt{--help} command line option.


\subsection{The \litterbox Web Interface}

\begin{figure}[t]
  \centering
  \includegraphics[width=\columnwidth]{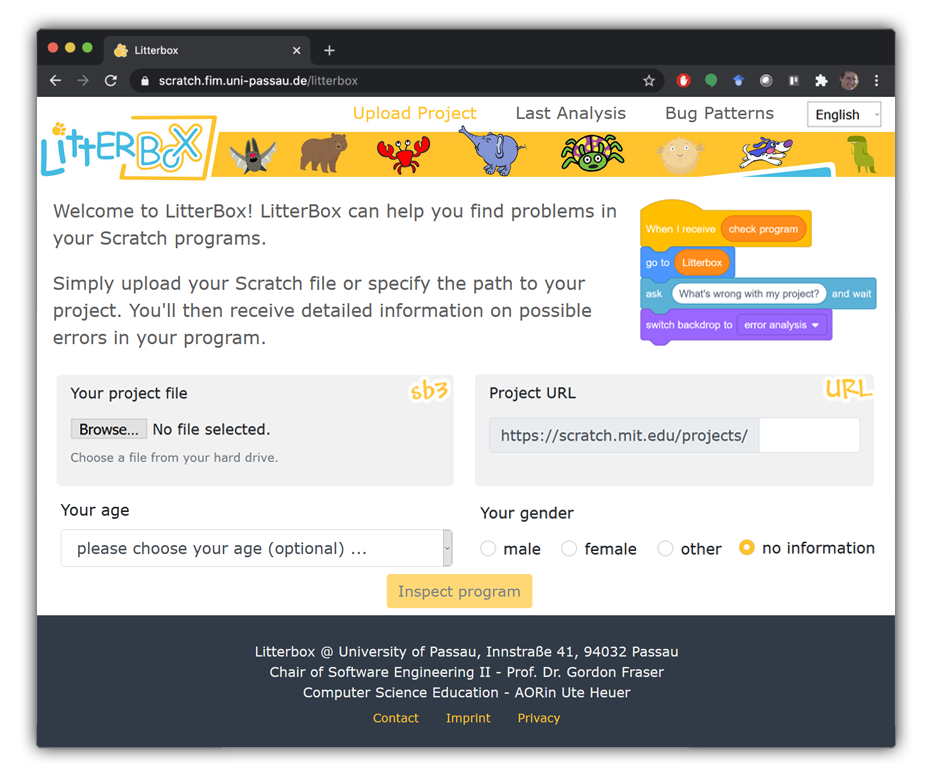}
  \caption{\label{fig:website-upload}Projects can be checked by uploading a file or entering the ID of a shared project.}
\end{figure}
\begin{figure}[t]
  \centering
  \includegraphics[width=\columnwidth]{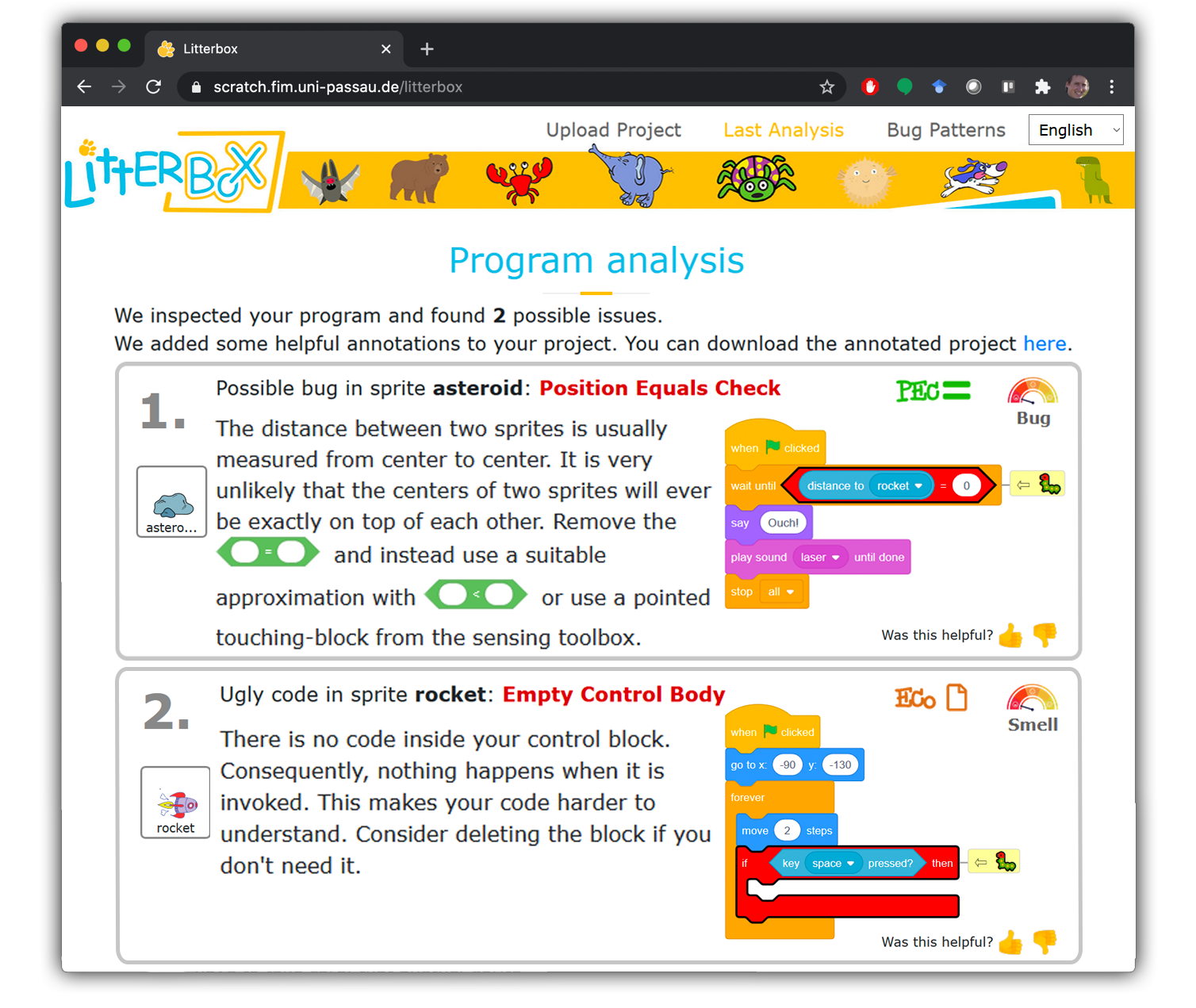}
\caption{\label{fig:website-result}The result of the analysis is a list of issues and hints.}
\end{figure}

For easier usage by learners \litterbox can also be accessed via a web interface. We provide a publicly accessible version of this at 
	{\url{https://scratch-litterbox.org}}.

The web interface supports multiple languages (currently English, German, and Spanish). 
Users can upload a \texttt{.sb3} file or simply provide the ID of a publicly shared \scratch project (see Figure~\ref{fig:website-upload}). 
After the analysis is finished, a list of issues found is shown (Figure~\ref{fig:website-result}). For each issue, we display (1) the sprite which contains the affected script, (2) the block-code representation of the erroneous script, with the affected block highlighted in red with a worm attached to it, and (3) a textual description of the problem and possible remedies.


\begin{figure}[t!]
\begin{lstlisting}[language=Java,frame=tb]
public class MissingAsk extends AbstractIssueFinder  {

    private List<Answer> answerBlocks = new ArrayList<>();
    private boolean askUsed = false;

    @Override
    public Set<Issue> check(Program program) {
        // ...
        program.accept(this);

        // Add an issue for each answer block 
        // if the program contains no ask-block
        if (!answerBlocks.isEmpty() && !askUsed) {
            for (Answer answer : answerBlocks) {
                addIssue(answer, answer.getMetadata());
            }
        }
        return issues;
    }

    @Override
    public void visit(AskAndWait node) {
        askUsed = true;
    }

    @Override
    public void visit(Answer node) {
        answerBlocks.add(node);
    }

    // ...
}
\end{lstlisting}
\caption[Example bug finder to check if the \emph{answer} block is used without an \emph{ask} block.]{Example bug finder to check if the \scratchblock{Answer} block is used without an \scratchblock{AskAndWait} block.}\label{lst:ask}
\end{figure}

\subsection{Extending \litterbox}

Extending \litterbox with new bug patterns or smell checks is easy, as
the internal structure allows extension by adding just a single class,
a so called \emph{finder}. As there is an abstract base class for
finders which implements an AST visitor, the concrete finders have to
override the existing \texttt{visit} methods only for the nodes that
are relevant for the new check.

For convenience, the AST offers different levels of abstraction, for
example, all reporter blocks implement one common interface
\texttt{Expression} and all stack blocks implement \texttt{Stmt}, so
that it is possible to handle all of them the same way without having
to implement one \texttt{visit} method for each expression or stack
block, respectively.

Figure~\ref{lst:ask} shows a simple bug finder which marks each usage
of an \scratchblock{answer} block as an issue if the program contains
no corresponding \scratchblock{ask} blocks.  To achieve this, the
finder simply needs to override the \texttt{visit} methods for the
\scratchblock{AskAndWait} and \scratchblock{Answer} blocks to keep
track of their usage, and afterwards can annotate all issues in the
\texttt{check} method.  \litterbox also makes more complex checks
possible, using the control flow graph or support for classical data
flow analysis.

Each finder needs to provide at least one hint message. It is possible
to customize the hint messages to take specific aspects of the issue
at hand into account, and provide more actionable help. Hint messages
are internationalized through the use of resource bundles.  Support
for new languages can simply be added by including new resource
bundles.


\section{Applications in Programming Education}

\subsection{Learners}

Previous research has already shown that code smells in \scratch
hamper the learning process of
students~\cite{hermans2016b}. \litterbox identifies not only code
smells, but also bug patterns, which are more severe issues than
smells, and thus likely have an even worse impact on learning.

\litterbox provides hints that consist of (1) an explanation of the
issue focusing on meaningful context and (2) a short description of
possible actions that might assist learners in solving the issue. This
helps learners segmenting the complex debugging and refactoring
process into manageable parts by locating potential issues one by one
(Figure~\ref{fig:website-result}). It may also ease reflection on
possible underlying misconceptions and foster resolving
them by targeting inappropriate concepts for repair.

Deploying Eccles's expectancy-value theory~\cite{eccles2009} on
students' debugging (and refactoring) processes yields the assumption
that learners are more likely to keep on debugging if they expect to
do well and they value debugging. As such, working with \litterbox
learners might be more motivated to succeed.

\subsection{Teachers}

\litterbox can also be used by teachers to get an overview of
potential bugs in their students' programs, in particular bugs that
might be signs of misconceptions or missing concepts. It can thus
support teachers in analyzing their learners' current knowledge and
skills, facilitating instructional adjustments repeatedly.  In
particular, \litterbox endorses teachers assigning different tasks
that are specifically set out to students differing in prior
performance, thus alleviating individualized instruction.

\litterbox supports students when fixing their broken programs without
needing extensive assistance of a human teacher while doing
so. Teachers are then able to spend more time with learners in need of
special support or extra attention. Note, however, that \litterbox
finds only generic issues that are problematic in any \scratch
program. Bugs that are specific to the task at hand require analysis
with respect to a
specification~\cite{stahlbauer2019testing,stahlbauer2020}.

\section{Empirical Results}

We investigated~\cite{fraedrich2020} a data set of \numnoremix
publicly shared \scratch projects (excluding remixes), and found a
total of \numbugs instances of the 25 bug patterns implemented in
\litterbox at the time of the study.
Not every instance of a bug pattern may lead to visible failure of the
program, and static analysis tools may in general produce false
positives. We therefore manually inspected a stratified random sample
of \numclassified bugs reported by \litterbox. In this sample, only
\numfalsepositives instances of bug patterns were revealed to be false
positives. Overall, this demonstrates that bug patterns occur
frequently in practice, and \litterbox helps to find them.



\section{Related Work}\label{sec:relatedWork}

The importance of analyzing \scratch programs was shown in previous studies~\cite{aivaloglou2016kids,hermans2016a,techapalokul2017a,robles2017software}, which found code smells to occur frequently in practice. Hermans and Aivaloglou~\cite{hermans2016b} determined empirically that
code smells may inhibit learning and make it harder for pupils to extend
\scratch programs with new functionalities. Techapalokul and
Tilevich~\cite{techapalokul2017a} also argue that code smells have negative
effects on the \scratch community, for example as projects with code smells are
less often remixed than those without smells.

The \hairball~\cite{boe2013} tool, which offers a basic Python API for
analysing \scratch (version 2) programs, was instrumental in enabling this
research on code smells, and other tools such as
\qualityhound~\cite{techapaloku2017b} followed for statically finding code
smells. \litterbox provides a generic program analysis framework for \scratch (version 3) programs. While it also incorporates basic code smells, it introduces the concept of bug patterns, the prevalence of which we demonstrated in an analysis of randomly sampled \scratch projects~\cite{fraedrich2020}. Whereas code smells identify low quality code that may lead to bugs in the future, patterns are indicative of bugs already present in the code. \litterbox offers various program analysis techniques, such as control- and data-flow analysis, to support the definition of further code smells, bug patterns, and also code metrics.

\litterbox also incorporates elaborate, multi-language hints to explain the issues found in the code. The web frontend in which these hints are displayed is inspired by \drscratch~\cite{moreno2015}, which analyzes \scratch projects with respect to the computational thinking concepts applied.

\section{Conclusions}

\scratch is tremendously popular with learners, teachers, and
researchers. There is, however, a need to provide support for all
these user groups. This is what \litterbox aims to achieve: \litterbox
is a static analysis tool that automatically detects bug patterns and
code smells. A convenient web-interface and helpful explanations
support learners and teachers alike, while the flexible command line
interface and a modular design support teachers and researchers.

In this paper we have summarized the features of \litterbox. By
providing \litterbox as open source, we hope to foster research on
analyzing \scratch programs as well as on \scratch-based programming
education.

To try out \litterbox, visit our web site:
\begin{center}
    \small{\texttt{\url{https://scratch-litterbox.org}}}
\end{center}

The source code of \litterbox is available at:
\begin{center}
    \small{\texttt{\url{https://github.com/se2p/LitterBox}}}
\end{center}

\section*{Acknowledgements}
This work is supported by DFG project FR 2955/3-1 ``Testing,
Debugging, and Repairing Blocks-based Programs'' and BMBF project
``primary::programming'' as part of the Qualit\"{a}tsoffensive
Lehrerbildung. We would like to thank Christoph Fr\"{a}drich, Miriam
M\"{u}nch, Gregorio Robles, Andreas Stahlbauer, and all other \litterbox contributors.


\balance
\bibliographystyle{IEEEtranS}
\bibliography{references}

\begin{thebibliography}{10}
\providecommand{\url}[1]{#1}
\csname url@samestyle\endcsname
\providecommand{\newblock}{\relax}
\providecommand{\bibinfo}[2]{#2}
\providecommand{\BIBentrySTDinterwordspacing}{\spaceskip=0pt\relax}
\providecommand{\BIBentryALTinterwordstretchfactor}{4}
\providecommand{\BIBentryALTinterwordspacing}{\spaceskip=\fontdimen2\font plus
\BIBentryALTinterwordstretchfactor\fontdimen3\font minus
  \fontdimen4\font\relax}
\providecommand{\BIBforeignlanguage}[2]{{%
\expandafter\ifx\csname l@#1\endcsname\relax
\typeout{** WARNING: IEEEtranS.bst: No hyphenation pattern has been}%
\typeout{** loaded for the language `#1'. Using the pattern for}%
\typeout{** the default language instead.}%
\else
\language=\csname l@#1\endcsname
\fi
#2}}
\providecommand{\BIBdecl}{\relax}
\BIBdecl

\bibitem{aivaloglou2016kids}
E.~Aivaloglou and F.~Hermans, ``{How kids code and how we know: An exploratory
  study on the Scratch repository},'' in \emph{Proceedings of the ACM
  Conference on International Computing Education Research}, 2016, pp. 53--61.

\bibitem{boe2013}
B.~Boe, C.~Hill, M.~Len, G.~Dreschler, P.~Conrad, and D.~Franklin, ``Hairball:
  Lint-inspired static analysis of scratch projects,'' 2013, pp. 215--220.

\bibitem{eccles2009}
J.~Eccles, ``{Who Am I and What Am I Going to Do With My Life? Personal and
  Collective Identities as Motivators of Action},'' \emph{Educational
  Psychologist}, vol.~44, no.~2, pp. 78--89, 2009.

\bibitem{fraedrich2020}
C.~Frädrich, F.~Obermüller, N.~Körber, U.~Heuer, and G.~Fraser, ``{Common
  Bugs in Scratch Programs},'' in \emph{Proceedings of the 2020 ACM Conference
  on Innovation and Technology in Computer Science Education}, ser. ITiCSE
  '20.\hskip 1em plus 0.5em minus 0.4em\relax ACM, 2020, pp. 89--95.

\bibitem{hermans2016b}
F.~Hermans and E.~Aivaloglou, ``{Do code smells hamper novice programming? A
  controlled experiment on Scratch programs},'' in \emph{2016 IEEE 24th
  International Conference on Program Comprehension (ICPC)}, 2016, pp. 1--10.

\bibitem{hermans2016a}
F.~Hermans, K.~T. Stolee, and D.~Hoepelman, ``{Smells in Block-Based
  Programming Languages},'' in \emph{2016 {{IEEE Symposium}} on {{Visual
  Languages}} and {{Human}}-{{Centric Computing}} ({{VL}}/{{HCC}})}.\hskip 1em
  plus 0.5em minus 0.4em\relax {IEEE}, 2016, pp. 68--72.

\bibitem{maloney2010}
J.~Maloney, M.~Resnick, N.~Rusk, B.~Silverman, and E.~Eastmond, ``{The Scratch
  Programming Language and Environment},'' \emph{ACM Transactions on Computing
  Education (TOCE)}, vol.~10, p.~16, 2010.

\bibitem{meerbaum2011habits}
O.~Meerbaum-Salant, M.~Armoni, and M.~Ben-Ari, ``{Habits of Programming in
  Scratch},'' in \emph{Proceedings of the Conference on Innovation and
  Technology in Computer Science Education}, 2011, pp. 168--172.

\bibitem{moreno2015}
J.~Moreno-León, G.~Robles, and M.~Román-González, ``Dr. scratch: Automatic
  analysis of scratch projects to assess and foster computational thinking,''
  \emph{RED-Revista de Educación a Distancia}, 2015.

\bibitem{scratch-for-all}
M.~Resnick, J.~Maloney, A.~Monroy-Hern\'{a}ndez, N.~Rusk, E.~Eastmond,
  K.~Brennan, A.~Millner, E.~Rosenbaum, J.~Silver, B.~Silverman, and Y.~Kafai,
  ``Scratch: Programming for all,'' \emph{Commun. ACM}, vol.~52, no.~11, p.
  60–67, 2009.

\bibitem{robles2017software}
G.~Robles, J.~Moreno-Le{\'o}n, E.~Aivaloglou, and F.~Hermans, ``{Software
  clones in Scratch projects: On the presence of copy-and-paste in
  computational thinking learning},'' in \emph{2017 IEEE 11th International
  Workshop on Software Clones (IWSC)}.\hskip 1em plus 0.5em minus 0.4em\relax
  IEEE, 2017, pp. 1--7.

\bibitem{stahlbauer2020}
A.~Stahlbauer, C.~Frädrich, and G.~Fraser, ``{Verified from Scratch: Program
  Analysis for Learners’ Programs},'' in \emph{In Proceedings of the
  International Conference on Automated Software Engineering (ASE)}.\hskip 1em
  plus 0.5em minus 0.4em\relax {IEEE}, 2020.

\bibitem{stahlbauer2019testing}
A.~Stahlbauer, M.~Kreis, and G.~Fraser, ``{Testing Scratch Programs
  Automatically},'' in \emph{Proceedings of the 2019 27th ACM Joint Meeting on
  European Software Engineering Conference and Symposium on the Foundations of
  Software Engineering}, 2019, pp. 165--175.

\bibitem{techapaloku2017b}
P.~Techapalokul and E.~Tilevich, ``Quality hound — an online code smell
  analyzer for scratch programs,'' in \emph{2017 IEEE Symposium on Visual
  Languages and Human-Centric Computing (VL/HCC)}, 2017, pp. 337--338.

\bibitem{techapalokul2017a}
------, ``{Understanding Recurring Quality Problems and Their Impact on Code
  Sharing in Block-Based Software},'' in \emph{2017 {{IEEE Symposium}} on
  {{Visual Languages}} and {{Human}}-{{Centric Computing}}
  ({{VL}}/{{HCC}})}.\hskip 1em plus 0.5em minus 0.4em\relax {IEEE}, 2017, pp.
  43--51.

\end{thebibliography}
 
\end{document}